\documentclass[]{fundam}
\usepackage{graphicx}
\usepackage[utf8x]{inputenc}
\usepackage[T1]{fontenc}

\usepackage{algorithmic}

\usepackage{amsmath,amssymb,amsfonts}
\usepackage{mathrsfs}
\usepackage{delarray}
\usepackage{array}
\usepackage{graphicx}
\usepackage{float}
\usepackage{tikz}
\DeclareMathOperator{\poly}{Poly}

\usepackage{booktabs}

\usepackage{cleveref}

\usepackage{url}

\usepackage{lineno}

\crefname{section}{§}{§§}
\Crefname{section}{§}{§§}

\newtheorem{thm}{Theorem}

\newtheorem{Cnjt}{Conjecture}

\DeclareMathOperator\ID{ID}
\def\CID{\textsc{Cid}}
\DeclareMathOperator\STL{ttl}
\DeclareMathOperator\STN{TTN}

\def\pr{\mathbb{P}}

\def\ELIMINATOR{\textsc{Eliminator}}
\def\NULL{\textsc{Null}}
\def\BL{\textsc{BL}}
\def\BCDL{\textsc{BcdL}}
\def\RNnoCD{\textsc{RNnoCD}}

\def\BEEP{\textsc{Beep}}

\def\TEST{\textsc{Test}} 

\def\ELIMINATED{\textsc{Eliminated}}

\def\status{\textsc{Status}}

\DeclareMathOperator\elll{l}
\def\prev{\textsc{Prev}}

\def\DETNAML{\textsc{DeterministicNaming}}

\def\RANDNAML{\textsc{RandomizedNaming}}

\def\CANDIDATE{\textsc{Active}}

\def\ALGOTEST{1}

\def\ALGODETLINIT{2}

\def\ALGOINITN{3}

\usepackage{pdfpages}

\def \beq{\begin{equation}}
\def \eeq{\end{equation}}
\def \be{\begin{eqnarray*}}
	\def \ee{\end{eqnarray*}}
\def \ben{\begin{eqnarray}}
\def \een{\end{eqnarray}}

\newtheorem{Rem}{Remark}

\newcommand*{\Endproof}{\hfill\ensuremath{\square}}%

\usepackage[linesnumbered]{algorithm2e}

\usepackage{authblk}

\begin{document}
%
\setcounter{page}{1}
\publyear{2021}
\papernumber{0001}
\volume{178}
\issue{1}
%

\title{Energy-Efficient Naming in Beeping Networks}

\address{Address for correspondence goes here}

\author{Ny Aina \textsc{Andriambolamalala}\\
	Universit\'e Paris Cit\'e, IRIF, CNRS, F-75013 Paris, France \\
	ny-aina.andriambolamalala@irif.fr
	\and Vlady \textsc{Ravelomanana}\\
	Universit\'e Paris Cit\'e, IRIF, CNRS, F-75013 Paris, France \\
	vlad@irif.fr } 

\maketitle

\runninghead{N. A. Andriambolamalala, V. Ravelomanana}{Energy-efficient Naming in Beeping Networks}

\begin{abstract}

A single-hop beeping network is a distributed communication model in which all stations can communicate with one another by transmitting only $1-bit$ messages, called beeps. This paper focuses on resolving the distributed computing area's two fundamental problems: naming and counting problems. We are particularly interested in optimizing the energy complexity and the running time of algorithms to resolve these problems. Our contribution is to design randomized algorithms with an optimal running time of $O(n \log n)$ and an energy complexity of $O(\log^2 n)$ for both the naming and counting problems on single-hop beeping networks of $n$ stations.

\end{abstract}

\setcounter{footnote}{0}
\section{Introduction}
Introduced by Cornejo and Kuhn in 2010~\cite{CornejoKuhn}, the beeping model makes minor demands on the network devices, which only need to be capable of performing carrier-sensing to differentiate between silence and the presence of a jamming signal (considered to be a $1-bit$ message or one beep) on the network. Such devices have an unbounded local power computation~\cite{chang2017exponential}. The authors of~\cite{CornejoKuhn} noted that carrier-sensing can typically be performed much more reliably and requires significantly less energy and resources than the message-sending model's devices. The minimization of such energy consumption per node arises, as the nodes are battery-powered most of the time. Since sending or receiving messages costs more energy than performing internal computations, a distributed algorithm's energy consumption is measured by the maximal over the waking time slot's number of all nodes~\cite{chang2017exponential,OLARIU,lavault2007quasi,Polonais,kardas2013energy,vlady2016time} (a node is awake when it beeps or listens to the network). It is more realistic to consider the nodes with no prior information regarding the network's topology and no identifiers (all nodes are initially indistinguishable).
Symmetry breaking protocols are important tools that allow the design of more elaborate algorithms. Thus, researchers designed various symmetry breaking protocols, such as leader election~\cite{chang2017exponential,kutten2013sublinear,jurdzinski2002efficient,LE_SODA13,GhaffariLynchSastry,lavault2007quasi,ramanathan2007randomized,UNIform_LE,kardas2013energy}, maximal independent set~\cite{AfekETAL,scott2013feedback}, and naming protocols~\cite{nakano1995optimal,hayashi1999randomized,OLARIU,bordim1999energy,chlebus2017naming}.
In this paper, we consider the naming problem of the single-hop\footnote{The underlying graph of the network is complete.} beeping networks of unknown size $n$, which assigns a unique label $\elll(s) \in\{1,\,2,\cdots,\, n\}$ to each node $s.$ We design an energy-efficient randomized naming algorithm succeeding in $O(n\log n)$ time slots with high probability \footnote{An event $\varepsilon_n$ occurs with
	high probability if $\pr[\varepsilon_n]\geq 1-\frac{1}{n^c}$ for some constant $c>0$, \textit{i.e.}, the probability of error is inversely proportional to a polynomial in $n.$} (\textit{w.h.p.} for short) on such a model, 
with $O(\log^2 n)$ energy complexity. First, in Section~\ref{detnaminalgo}, we present a deterministic algorithm, naming some $M$ nodes (with $M\leq n$) in $O(M\log n)$ time slots with $O(M\log n)$ energy complexity. Then,
in Section~\ref{namingNknown}, we address the naming problem for $n$ nodes when the number of nodes, $n$, is unknown.
The algorithm developed to solve the naming problem is then adapted to address the counting problem,
which involves allowing each node to learn the total number of the nodes.
In Section~\ref{LB}, we conjecture a lower bound of $\Omega(\log n)$ on the beeping network's naming algorithm's energy complexity. Finally, in Section~\ref{maple}, we present some Maple Simulation results that illustrate our work.

\subsection{Considered models and complexities measure} 
\subsubsection{Beeping network model}
\begin{figure}[H]
	\includegraphics[width=\linewidth]{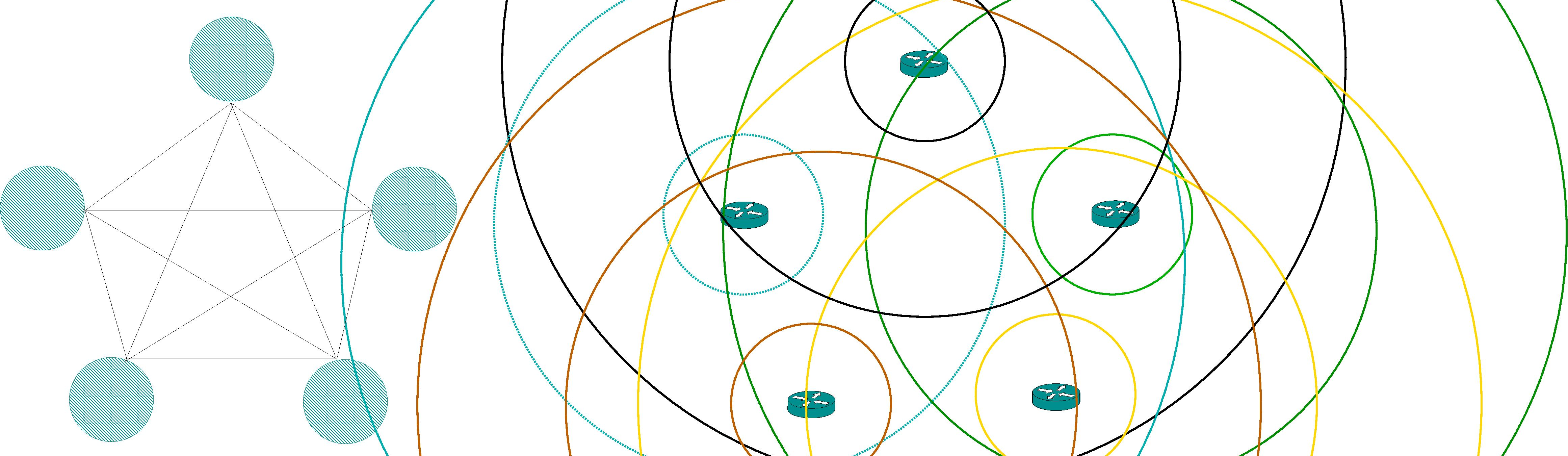}
\caption{}
\label{figmodel}
\end{figure}

In single-hop beeping networks, the nodes communicate with one another during discrete time slots via a shared beeping channel. Each node has access to a global clock. 
As presented in Figure~\ref{figmodel}, single-hop beeping networks can be used for modelling an ad-hoc network, in which all nodes are in one another's communication range. The nodes can send $1-bit$ messages and perform carrier sensing to detect any transmission. 
\noindent
At each time slot, a node independently determines whether
to transmit a beep, listen to the network, or remain idle (asleep). Only the listening nodes can receive the common channel's state : namely, \BEEP~if at least one node is transmitting or \NULL~if no node transmits.
This model is also called the $\BL$ or the beep listen model. 
\subsubsection{Complexities measure}
\begin{definition}[Time complexity]
  The time complexity of an algorithm is defined as the maximum number of time slots required for all nodes in the network to complete the algorithm.
\end{definition}

There are several definitions of energy complexity for a distributed algorithm, depending on the specific model of transmission and reception energy cost considered. In this paper, we provide a formal definition of energy complexity as follows~\cite{chlebus2017naming,chang2018energy,chang2019exponential}. 
\begin{definition}[Energy complexity]
  A node is considered active if it is transmitting or listening to the network. In this paper, the energy complexity of a distributed algorithm is defined as the maximum number of active time slots per node, over all nodes
  participating in the algorithm.
\end{definition}

\begin{Rem}
	Some papers consider only transmissions due to the negligible cost of message receptions to measure a distributed algorithm's energy complexity~\cite{sivalingam1997low}. In this paper, we consider both the transmission and the message reception.
\end{Rem}

\subsection{Related works and new results}  
As the naming problem is a fundamental distributed computing problem~\cite{nakano1995optimal}, many results exist for this problem. Let us first consider the simplest model, namely the single-hop network, in which the network's underlying graph is complete. In~\cite{hayashi1999randomized}, Hayashi, Nakano and Olariu presented a randomized naming protocol with $O(n)$ running time for radio networks with collision detection (or RNCD for short), in which the nodes transmit messages of size $O(\log n).$ Note that on RNCD, the nodes can only transmit or listen to the network during each time slot, and the listening nodes receive only single transmissions. If more than one node transmits simultaneously, a collision is detected by all the listening nodes. Later, Bordim, Cui, Hayashi, Nakano and Olariu~\cite{bordim1999energy} presented a naming algorithm for RNCD, terminating \textit{w.h.p.} in $O(n)$ time slots with $O(\log n)$ energy complexity. In~\cite{nakano2000energy}, for radio networks with no collision detection (\RNnoCD~for short), Nakano and Olariu designed a naming protocol, terminating in $O(n)$ time slots \textit{w.h.p.} with $O(\log\log n )$ energy complexity. On \RNnoCD, all nodes can transmit messages of size $O(\log n)$ during each time slot. Meanwhile, for RNCD, only the messages transmitted alone are received, but the devices can detect no collision. As the naming and counting problems are closely related to the network's size approximation problem, Jurdzi{\'n}ski, Kuty{\l}owski and Zatopia{\'n}ski presented an energy-efficient algorithm to address this problem with $o(\log\log n)$ energy complexity~\cite{Polonais}.

Results on the beeping model appeared recently when Chlebus, De Marco and Talo~\cite{chlebus2017naming} presented their naming algorithm terminating in $O(n \log n)$ time slots \textit{w.h.p.} for the $\BL$ model. Moreover, they also provided a $\Omega(n\log n)$ lower bound on such an algorithm's time complexity. Meanwhile, Casteigts, M{\'e}tivier, Robson and Zemmari~\cite{casteigts2016counting} presented a counting algorithm for the $\BCDL$\footnote{The transmitters can detect collisions on the $\BCDL$ model or Beep with Collision Detection Listen~\cite{AfekETAL,scott2013feedback}.
} model terminating in $O(n)$ time slots \textit{w.h.p.} They noticed that adapting their algorithm to work on the $\BL$ model would result in a logarithmic slowdown regarding its time complexity.  
\noindent
The following Table~\ref{tab1} compares our results to the existing results of the single-hop network model. In Table~\ref{tab1}, as $n$ is unknown, $N$ represents a polynomial upper bound on $n$ ($N = \poly(n)$\footnote{In this paper, we define $\poly(n)= n^\alpha$ for some constant $\alpha >1.$}), which is computed by the nodes using the approximate counting algorithm described in~\cite{brandes2016approximating}.

\begin{table}
\caption{Comparison between some existing results and ours.}
\label{tab1}
	\hspace{0cm}\begin{tabular}{cccc}
		\toprule
		Problem &  Time  & Energy & Probability of\\
		and model& complexity & complexity& success\\ 
				\toprule
		\multicolumn{4}{c}{Existing results } \\
		\toprule
		Randomized naming	  & & &\\		
		on single-hop \RNnoCD~\cite{nakano2000energy}&$O(n)$&$O(\log\log n)$& $1-O\left(\frac{1}{n}\right)$\\
		$n$ known&&&\\
		\hline
		
		Randomized naming&  & &\\
		on single-hop beeping networks~\cite{chlebus2017naming} &$O(n\log n)$&Not applicable& $1-O\left(\frac{1}{n}\right)$\\
		 ($\BL$), $n$ known& & &\\
		\hline
		Randomized counting	& & &\\		
		on single-hop beeping networks~\cite{casteigts2016counting}&$O(n)$&Not applicable& $1-O\left(\frac{1}{n}\right)$\\
		($\BCDL$)&&&\\
		\toprule
		\multicolumn{4}{c}{\textbf{Our results} } \\
		\toprule
		\hline
		Randomized naming &&&\\
		on single-hop $\BL$ &$O(n\log n)$ & $O(\log^2 n)$&$1-O\left(n^{-c}\right)$\\
		Indistinguishable nodes, & & & for some $c>0$ \\
		$n$ unknown, Theorem~\ref{thm:randomunknown}&&&\\
				\hline
		Randomized&&&\\
		counting on single-hop $\BL$&$O(n\log n)$&$O(\log^2 n)$& $1-O\left(n^{-c}\right)$\\
		Indistinguishable nodes, &&&\\
		
		$n$ unknown, Theorem~\ref{thm:counting}&  & &for some $c>0$\\

		\bottomrule
			
	\end{tabular}

\end{table}

\noindent The more realistic model, in which the network's underlying graph is an arbitrary connected graph (called the multi-hop network model), gained importance as a research subject~\cite{perkins2001ad}. Czumaj and Davies~\cite{czumaj2019communicating} designed a deterministic depth-first search (DFS~for short) algorithm that initialized such a network in $O(n\log n)$ time slots. \cite{ravelomanana2005randomized} analyses the naming protocol on such a multi-hop case, but it is restricted to a set of nodes randomly located in a square area. 


\section{New approach: naming $M$ nodes in a deterministic manner}\label{detnaminalgo}
\subsection{A first energy-inefficient naming algorithm}
For the sake of simplicity without loss of generality, for the remainder of this section, let us suppose that each node $s$ knows a common upper bound $N$ on $n$ such that $N=\poly(n)$ and has a unique identifier denoted $\ID_s \in \{1,\,2,\,\cdots, \, N\}.$ We note that this assumption does not limit the generality of our results. 

Our new approach employs a known method representing the $\ID$s by their binary encoding and transmitting the obtained bits one at a time in reverse order~\cite{jacquet2013novel}. 
\begin{definition}[$\CID_s$]
 $\CID_s$ is the binary representation of a given $\ID_s.$ We have $\CID_s =
 \left[\CID_s[0]\CID_s[1]\cdots\CID_s[\lceil\log_2 N\rceil]\right]$ such that $\CID_s[i] \in \{0,1\}.$
\end{definition}


Let $M$ ($M \leq n$) represent the number of nodes participating in a specific stage or phase of the naming algorithm.
Each of the $M$ nodes first computes its $\CID_s$ and sends it, bit by bit, over $\lceil 2\log_2 N +2\rceil = O(\log n)$ time slots to check whether it has the largest $\ID$ of all nodes. The node with the largest $\ID$ will take the next available label. During each time slot, a node $s$ that detects that another node $v$ has a higher $\ID_v$ than $\ID_s$ is eliminated (\textit{i.e.}, it is no longer a candidate to take the next label). Thus, after $\lceil 2\log_2 N +2\rceil $ time slots, the unique remaining candidate node holding the largest $\ID$ gets labelled.

 \begin{definition}[Season]
	A season consists of $\lceil2\log_2 N+2\rceil$ time slots. 
\end{definition}

 In what follows, we consider our naming algorithm as successive seasons denoted $S_1, \, S_2, \,
 \cdots, \, S_j,\,\cdots ,\, S_M$ (note that the nodes do not necessarily know $M$). During season $S_j$, each node $s$ sends its $\CID_s$ bit by bit over $\lceil2\log_2 N+2\rceil$
time slots $t_0,\, t_1,\, \cdots,\, t_{2\lceil\log_2 N \rceil+1}.$ To do so, for $i = 0,\,1,\,\cdots,\,\lceil\log_2 N+1\rceil$, during each couple of time slots $\{t_{2i},\ t_{2i+1} = t_{2i}+1\}$, each node executes the following defined \TEST($2i$) protocol. \TEST($2i$) accepts an even time slot number $2i$ as a parameter and outputs a status $\in \{\ELIMINATED,$ $ \CANDIDATE,\ELIMINATOR\}$ at the end of $t_{2i+1}.$ By executing \TEST($2i$), a node $s$ can take one status according to the following rules: 

\begin{itemize}
	\item If $s$ has $\CID_s[i] = 0$ and detects that at least one node $v$ ($v\neq s$) has $\CID_v[i] = 1$, it becomes \ELIMINATED~at the end of $\{t_{2i},\, t_{2i+1}\}.$ 
	\item This latter node $v$ becomes an \ELIMINATOR.  
	\item If $s$ neither is eliminated nor has eliminated some other node during  $t_{2i},\, t_{2i+1}$, it stays \CANDIDATE. 
\end{itemize}
\noindent
Let us now define this \TEST($2i$) protocol.\\

\noindent
\underline{\textbf{Algorithm~\ALGOTEST.}}~ \TEST$(2i)$ at any node $s.$\\
\framebox[1.\width]{
	\begin{algorithm}[H]
		\DontPrintSemicolon
		\SetKwInOut{Input}{Input}
		\SetKwInOut{Output}{Output}
		\Indm
		\Input{The node $s$ has a unique code-word denoted $\CID_s$ and receives the time slot's number $2i$
                           as a parameter.}
		\Output{$\status(s)\in\{\CANDIDATE,\,\ELIMINATED,\,\ELIMINATOR\}$}
		\BlankLine
		\Indp 
		\eIf{$\CID_s[i]=0$}{
			$s$ beeps at $t_{2i}$ and listens to the network at $t_{2i+1} = t_{2i}+1$\\
			\eIf{$s$ {\upshape hears a beep at} $t_{2i+1}$}{
				$s$ sets $\status(s)\leftarrow$ \ELIMINATED.
			}{$s$ sets $\status(s)\leftarrow$ \CANDIDATE.}
		}{
			$s$ listens to the network at $t_{2i}$ and beeps at $t_{2i+1}$\\
			\eIf{$s$ {\upshape hears a beep at} $t_{2i}$}{
				$s$ sets $\status(s)\leftarrow$ \ELIMINATOR.	
			}{$s$ sets $\status(s)\leftarrow$ \CANDIDATE.}
		}
		
	\end{algorithm}
}\par

\bigbreak

\paragraph{Our deterministic naming algorithm's description for $M$ nodes.}
Consider a set of $M$ participating nodes during the following protocol.
At each even time slot $t_{2i}$, where $i = 0, \,1,\, 2, \cdots,\, \log_2 N$,
in each season $S_j$, where $j \in \{1,\, 2,\, \cdots,\, M\}$, each node $s$ executes the \TEST($2i$) function.
If at least one node $v$ has $\CID_v[i] = 1$, then each node $s$ with $\CID_s[i] = 0$
is eliminated until the beginning of the next season $S_{j+1}$. At the end of season $S_j$,
the last remaining node is assigned the label $j$. By running this process for each $j \in \{1,\, 2,\, \cdots, M\}$, we
 obtain the following.
\begin{lemma}\label{lemma:MLOGN}
The naming algorithm described above terminates
 terminates in $O(M \log n)$ time slots and assigns a unique label in $\{1,\,\cdots,\, M\}$ to each of the $M$ nodes.
 \end{lemma}
\noindent
\subsection{Energy optimization principle}\label{(iii)} The previously presented algorithm is not energy-efficient as all nodes wake up (to beep or listen to the network) during the $O(M\log n)$ time slots. To improve this energy consumption, we will subdivide the set of participants into small groups and pass the last identifier of a group to the next group to keep an order in the successive numberings. To do this, we need a deterministic algorithm capable of numbering the members of a group without knowing the exact cardinality of this group. 
To do this, we need to define several sub-procedures and definitions.

\begin{definition}[\textbf{Time To Listen or $\STL(s)$} for short]\label{stl} 
	 
	 $\STL(s)$ is an even time slot $t_{2i}, \, \mbox{ for }i=0,\,1,\, \cdots, \,\lceil\log_2 N \rceil,$ of any even season $S_{2j-1}, \, \mbox{for } j = 1,\,2,\,\cdots,\, M,$ such that the node $s$ has been eliminated during the couple of time slots $(t_{2i}, \, t_{2i}+1)$, that is, the node $s$ has $\CID_s[i] = 0$, and some unlabeled node $v$ has $\CID_v[i]=1$, so $s$ sets $t_{2i}$ as its $\STL(s).$
	 \[
	 \STL(s) = \max_{i=0,\,1,\,\cdots} \{t_{2i} \mbox{ such that } \CID_s[i] = 0 \mbox{ and } \exists \mbox{ an unlabeled node } v (v \neq s) \mbox{ such that }  \CID_v[i] = 1 \}.
	 \]

\end{definition}

\noindent
For $i=0,\,1,\, \cdots, \,\lceil\log_2 N \rceil$ and $j = 1,\,2,\,\cdots,\, M$, a node $s$ must wake up and execute \TEST$(2i)$ at $t_{2i} = \STL(s)$ during the even seasons $S_{2},\, S_{4},\, \cdots,\, S_{2k}$ where $S_{2k}$ is the even season during which $s$ is no longer eliminated.

\begin{definition}[\textbf{Time To Notify or $\STN(s)$}]\label{stn}
	$\STN(s)$ is a set of even time slots $\{t_{2i}\}, \,\mbox{for } i=0,\,1,\, \cdots, \,\lceil\log_2 N \rceil, $
        of any odd season $S_{2j-1}, \, \mbox{for } j = 1,\,2,\,\cdots,\, M,$ such that
        the node $s$ has eliminated some other nodes during the couple of time slots $(t_{2i}, \, t_{2i}+1).$
        More precisely, if during any season $S_j, \, j = 1,\,2,\,\cdots,\, M,$
        the node $s$ has $\CID_s[i] = 1$ and some remaining unlabeled node $w$ has $\CID_w[i] = 0$,
        then $s$ stores $t_{2i}$ in $\STN(s).$
	\[
	\STN(s) = \{t_{2i}\}_{i=0,\,1,\,\cdots} \mbox{ such that } \CID_s[i] = 1 \mbox{ and } \exists \mbox{ an unlabeled node } w (w \neq s ) \mbox{ such that }  \CID_w[i] = 0.
	\]
\end{definition}

\noindent
Let $S_l$ be the season during which the node $s$ gets labelled. For $i=0,\,1,\, \cdots, \,\lceil\log_2 N \rceil, \,  j = 1,\,2,\,\cdots,\, M$, during the seasons $S_{j+1},\, S_{j+2},\, \cdots,\, S_l$, the node $s$ must execute \TEST$(2i)$ at each time slot $t_{2i}$, saved until season $S_j$ in its $\STN(s)$, to notify the other nodes still eliminated during these seasons.  

\begin{Rem}
For $i=0,\,1,\, \cdots, \,\lceil\log_2 N \rceil, \,$ when a node $s$ adds $t_{2i}$ into its $\STN(s)$, it no longer has
an active neighbor $w$ that holds $\CID_w[k] = 1, k>i.$ Thus, $s$ sets its $\STL(s)$ to a specific value (say $-2$
as a sentinel value) which corresponds to none of the even time slots $t_{2i} (i \in \{0,\,\cdots,\, \lceil \log_2 N \rceil\})$.
\end{Rem} 

In what follows, we use these two new definitions to design our deterministic naming algorithm.

\subsection{Description of our deterministic  naming algorithm}

\paragraph{Initial states of the nodes:}Before executing the following defined algorithm, each node is sleeping (its radio is switched off).  Each node $s$ initializes its local variables $\STL(s)$ $\status(s)$ to \NULL~and $\STN(s)$ to the empty set $\{\}.$  \\

\noindent
Each node can wake up at any couple of time slots $(t_{2i},\, t_{2i+1}),\, i=0,\,1,\, \cdots, \,\lceil\log_2 N \rceil,$ of any season $S_j,\,  j = 1,\,2,\,\cdots,\, M,$ by calling the \TEST$(2i)$ protocol. The node returns to the sleeping state after each call of \TEST$(2i).$ Each node $s$ with the \ELIMINATED~status sleeps (its radio is switched off and it cannot execute \TEST$(2i)$) until the next season, when it resets its status.

\paragraph{High-level description of the algorithm :} First, at time slot $t_{0}$ of season $S_1$, each node $s$ sets $\status(s)\leftarrow$\TEST$(0)$ (line $6$ of Algorithm~\ALGODETLINIT). The \ELIMINATED~nodes at $t_0$ record $0$ as their $\STL(s)$ (line $13$ of Algorithm~\ALGODETLINIT). Meanwhile, each node $v$ with the $\ELIMINATOR$ status adds $0$ into its $\STN(v)$ and sets its $\status(v)$ to \CANDIDATE~(line $16$ of Algorithm~\ALGODETLINIT). Afterwards, for each time slot $t_{2i},\, i=0,\, \cdots,\, \lceil\log_2 N\rceil$ of each season $S_j,\, j=1,\, \cdots,\, M$, each node $s$ executes \TEST$(2i)$ only if one of the following is satisfied (lines $5-7$ of Algorithm~\ALGODETLINIT) :
\begin{itemize}
	\item $\status(s) = \NULL$ ($s$ starts a new season), and $t_{2i}\in\{\STN(s) \bigcup \{\STL(s)\} \}$,
	\item $\status(s) = \CANDIDATE$ \textit{i.e.}, at the time slot $t_{2i-2}$, the node $s$ executed \TEST$(2i-2)$, which returned \CANDIDATE~or \ELIMINATOR. Note that at line $16$ of Algorithm~\ALGODETLINIT, all \ELIMINATOR~nodes become \CANDIDATE. 
\end{itemize}

\bigbreak\noindent
Each such execution of \TEST$(2i)$ works as described in the time slot $t_0$ : If \TEST$(2i)$ returns \ELIMINATED, the node $s$ records $t_{2i}$ as its $\STL(s)$ and sets its $\status(s)$ to \ELIMINATED~(line $4$ of Algorithm~\ALGOTEST~and line $13$ of Algorithm~\ALGODETLINIT). Otherwise, if \TEST$(2i)$ returns \ELIMINATOR, $s$ adds $t_{2i}$ into its $\STN(s)$ and sets its $\status(s)$ to \CANDIDATE~(line $11$ of Algorithm~\ALGOTEST~and line $16$ of Algorithm~\ALGODETLINIT). Finally, if \TEST$(2i)$ returns \CANDIDATE, the node $s$ sets it as its $\status(s)$~(lines $6$ and $13$ of Algorithm~\ALGOTEST~and line $10$ of Algorithm~\ALGODETLINIT). At the end of each season, the remaining unique \CANDIDATE~node takes the next available label, and before proceeding to the next season, each node $s$ resets its $\status(s).$ \\

\noindent
For better comprehension, in what follows, we illustrate one example of the \DETNAML($N$) algorithm's execution using a binary tree, as done in~\cite{fuchs2016dependence}. By doing so, we count the waking time slot's number of a given node (red or black node in the following figures).

\subsection{Representing the \DETNAML($N$) algorithm's execution with a binary tree}\label{ApB}

\paragraph{Legends for the following figures:} 
\begin{itemize}
	\item[Shapes] The hexagonal shapes above each node of the tree represent $\STL(s)$, the squares represent $\STN(s)$, and the circles represent the other waking time slots. 
	\item[Numbers] The number inside these shapes defines the season during which the node $s$ wakes up at a time slot. Meanwhile, the numbers outside the shapes represent the sleeping time slots (when the node does not execute \TEST$(2i)$). Finally, the numbers inside the tree's nodes represent each even time slot of the algorithm's execution. 
	
\end{itemize}

\begin{figure}[H]
	\centering
	
		\includegraphics[height=4cm]{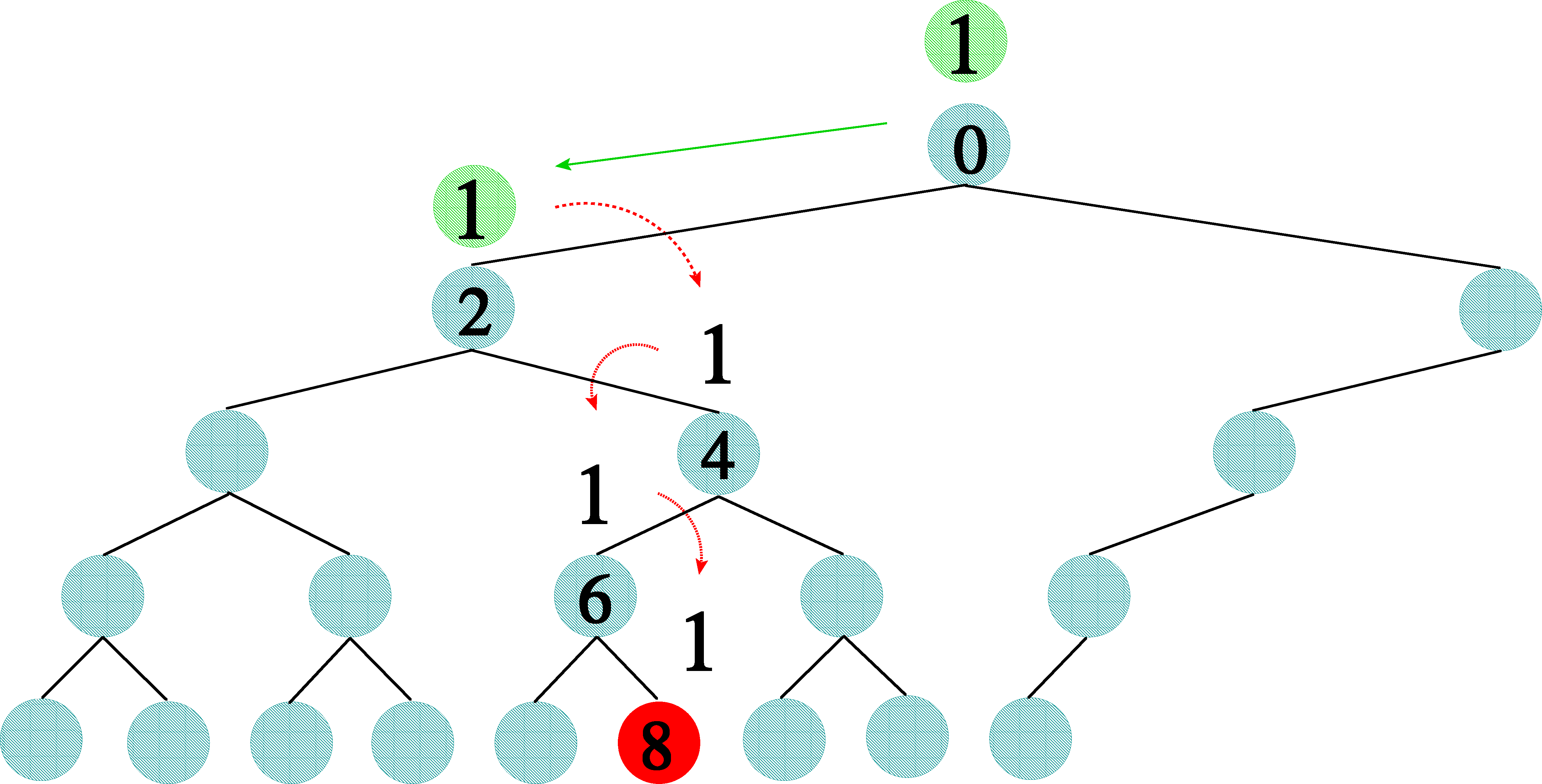}
		\caption{One path of the tree represents the $\CID_s$ of a node $s$, and one edge represents one bit of this $\CID_s$ : Left edges correspond to bit $1$, and right edges correspond to bit $0.$ For example, the considered red or black node $s$ has $\CID_s = 1010.$ Each such tree's node depicts each even time slot $t_{2i}$ of the algorithm's seasons, while the leaves represent the stations executing an algorithm. }	
	\label{fig:0}
\end{figure}

\begin{figure}[H]
	\centering
	\begin{minipage}{0.45\linewidth}	
		\includegraphics[height=2.9cm]{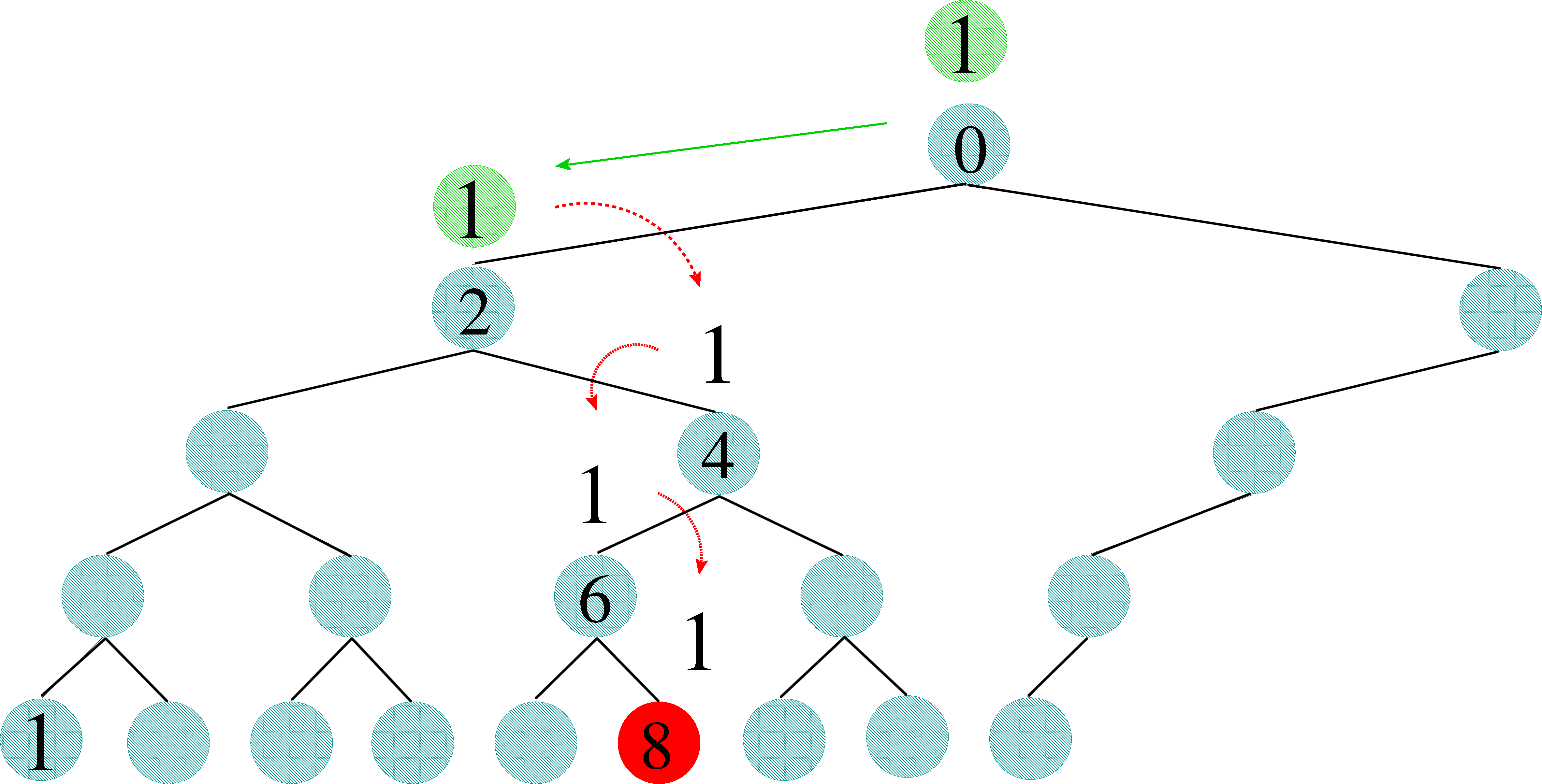}\\
		{{During season $S_1$, the node $s$ wakes up once at time slot $t_0$, adds $0$ into its $\STN(s)$, and wakes up and gets eliminated at $t_2.$ It sets $t_2$ as its $\STL(s)$ and sleeps until the beginning of season $S_2.$ Node~$1$ gets labelled at the end of season $S_1.$}}\\
	\end{minipage}	
	\hspace{1cm}\begin{minipage}{0.45\linewidth}	
		\includegraphics[height=2.9cm]{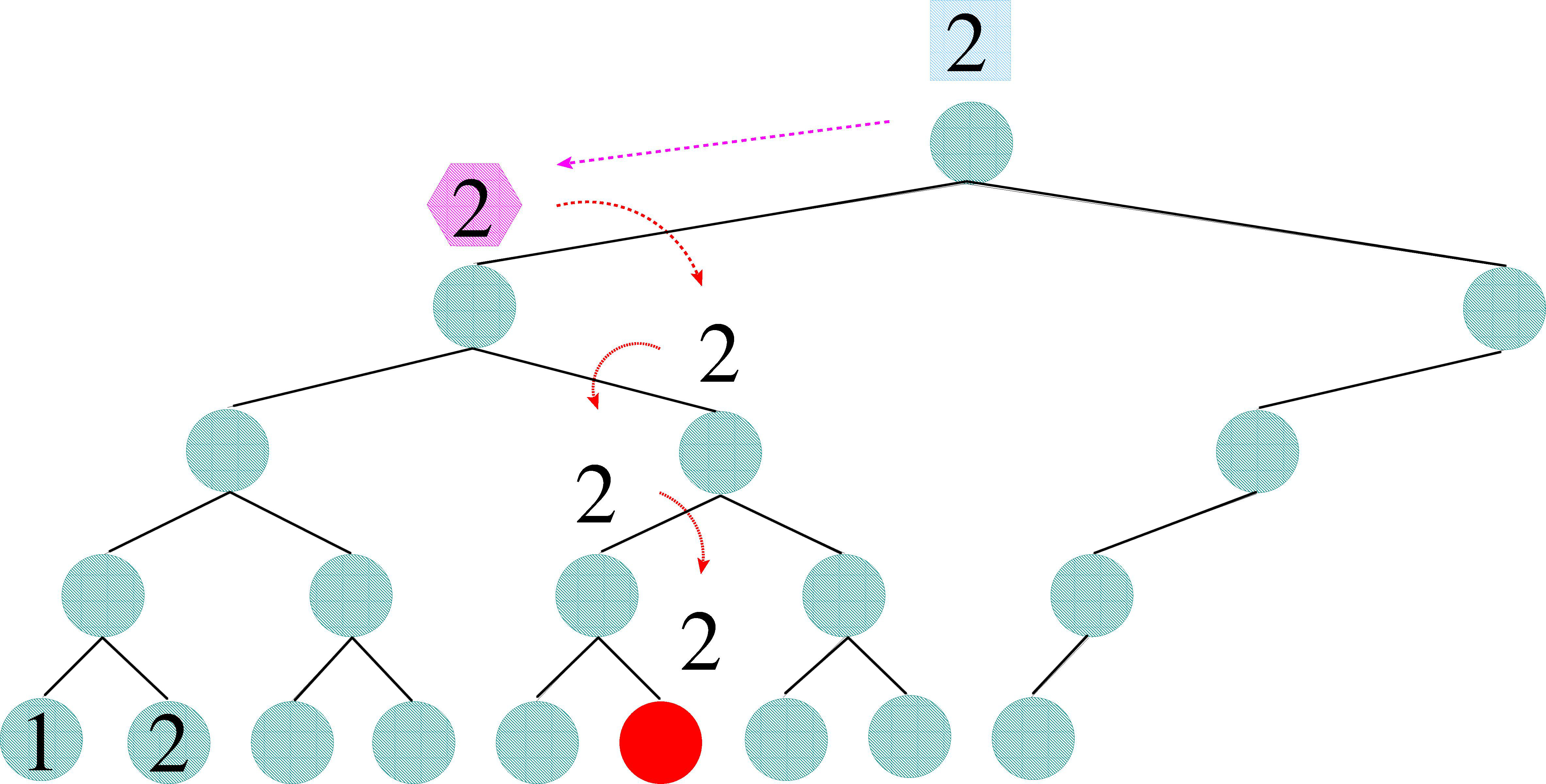}\\
		{{During $S_2$, $s$ wakes up at $t_0 \in \STN(s)$, sleeps, and wakes up at $t_2 \in \STL(s).$ Then, it gets eliminated at $t_2$, and sleeps until $S_3$ begins. Node~$2$ gets labelled at the end of $S_2.$$\,$\\$\,$\\$\,$}}\\
	\end{minipage}	
	
\end{figure}
\noindent
Seasons $S_3$ and $S_4$ work exactly as season $S_2.$

\begin{figure}[H]
	\begin{minipage}{0.45\linewidth}		
		\includegraphics[height=2.9cm]{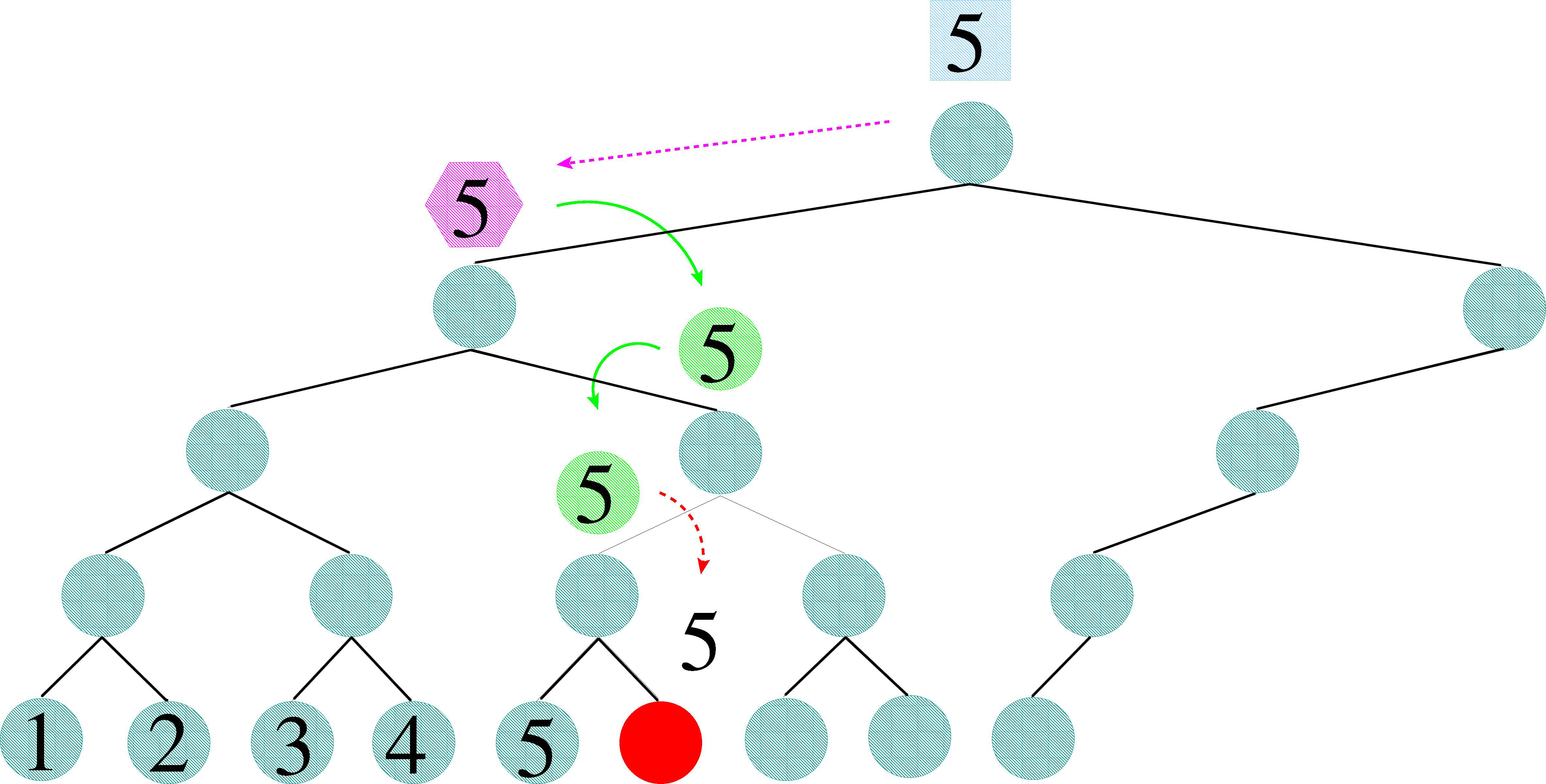}\\
		{{Then, during $S_5$, $s$ wakes up at $t_0 \in \STN(s)$, sleeps, and wakes up at $t_2 \in \STL(s).$ Since no more unlabeled node can eliminate $s$ at $t_2$, it wakes up at each time slot until $t_6$, when it gets eliminated. It adds $t_4$ into its $\STN(s)$, sets its $\STL(s)$ to $t_6$, and sleeps until $S_6$ begins. Node~$5$ gets labelled at the end of $S_5.$}}\\
	\end{minipage}	
	\hspace{1cm}\begin{minipage}{0.45\linewidth}		
		\includegraphics[height=2.9cm]{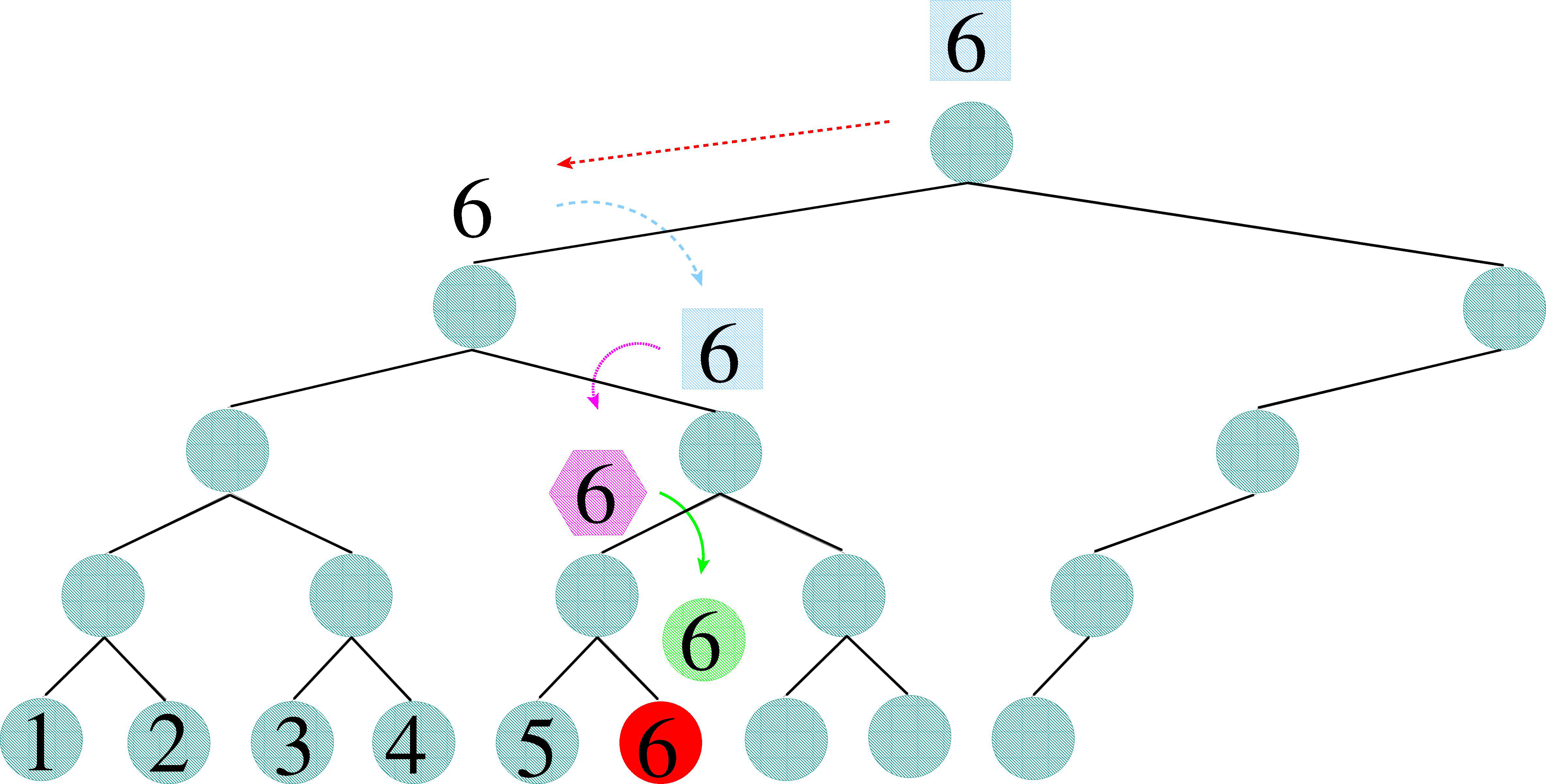}\\
		{{During $S_6$, $s$ wakes up at $t_0 \in \STN(s)$, sleeps, and wakes up at $t_4 \in \STN(s).$ It then sleeps and wakes up at $t_6\in \STL(s).$  Since no more unlabeled node can eliminate it, $s$ wakes up at $t_8$ and gets labelled.$\,$\\$\,$\\$\,$\\}}\\
	\end{minipage}
\end{figure}

\newpage
\noindent
\underline{\textbf{Algorithm~\ALGODETLINIT.}}~ \DETNAML($N$) at any node $s$\\
\framebox[1.\width]{
	\begin{algorithm}[H]
		\DontPrintSemicolon
		\SetKwInOut{Input}{Input}
		\SetKwInOut{Output}{Output}
		\Indm
		\Input{An upper bound $N$ on $n$ and a unique $\ID_s\in \{1,\,2,\,\cdots,\,  N\}.$}
		\Output{A unique label $\elll(s) \in \{1,\,2,\,\cdots,\, M\}.$}
		\BlankLine
		\Indp
		
		$s$ encodes its $\ID_s$ into a binary code-word $\CID_s \leftarrow \{0,1\}^{\lceil \log_2 N+1 \rceil}.$ \\
		$s$ sets $\elll(s) \leftarrow 0$, $\STL(s)\leftarrow \NULL$, $\STN(s)\leftarrow \{\}$, $j \leftarrow 1, \status(s)\leftarrow \NULL.$\\
		\While{$\elll(s) = 0$}{
			\For{$i \leftarrow 0$ {\upshape to } $\lceil \log_2 N \rceil$}{
				\If{$(\status(s) = \NULL$ and $(\STL(s) = t_{2i}$ or $t_{2i} \in\STN(s)$ or  $ j = 1))$ or  $(\status(s) = \CANDIDATE)$
				}{
					$s$ sets $\status(s) \leftarrow \TEST(2i)$ and sleeps.
				}
				
				\Switch{$\status(s)$}{
					\Case{$\CANDIDATE$}{
						$s$ removes $t_{2i}$ from its $\STN(s).$\\
						\If{$\CID_s[i]= 0$}{ resets $\STL(s).$ }
					}
					\Case{$\ELIMINATED$}{$s$ sets $t_{2i}$ as its $\STL(s).$}
					\Case{$\ELIMINATOR$}{
						$s$ adds $t_{2i}$ into its $\STN(s)$ and sets $\status(s) \leftarrow \CANDIDATE.$
					}
				}
				
			}
			
			\If{$\status(s) = \CANDIDATE$}{
				$s$ sets $\elll(s) \leftarrow j.$
			}	
			$s$ sets $j\leftarrow j + 1.$
		}
	\end{algorithm}
} \par

\begin{lemma}\label{Lemma:detlinit}
	In single-hop beeping networks of size $n$, there exists a deterministic algorithm that
        names $M$ distinct ($M \leq n$)
        nodes in $O(M\log n)$ time slots, with all nodes being awake the whole execution of the algorithm.
\end{lemma}
\noindent

\noindent
In what follows, let $W(s)$ be the total number of waking time slots of any node $s$ during Algorithm~\ALGODETLINIT's execution; let $W_{\STN}(s)$, $W_{\STL}(s)$ and $W_{o}(s)$ correspond to the $\STN(s)$ total waking time slots, the $\STL(s)$, and the other total waking time slots, respectively. Moreover, let $w, \, y, \, z$ be the nodes with the largest number of $\STL, \, \STN $, and other waking time slots, respectively. Thus, we have
\beq\label{eveilTotal}
W(s) = W_{\STN}(s) + W_{\STL}(s) + W_{o}(s)\leq W_{\STL}(w)+ W_{\STN}(y) +W_{o}(z) \, .
\eeq

\begin{figure}[H]
	\centering
	\includegraphics[height=4.5cm]{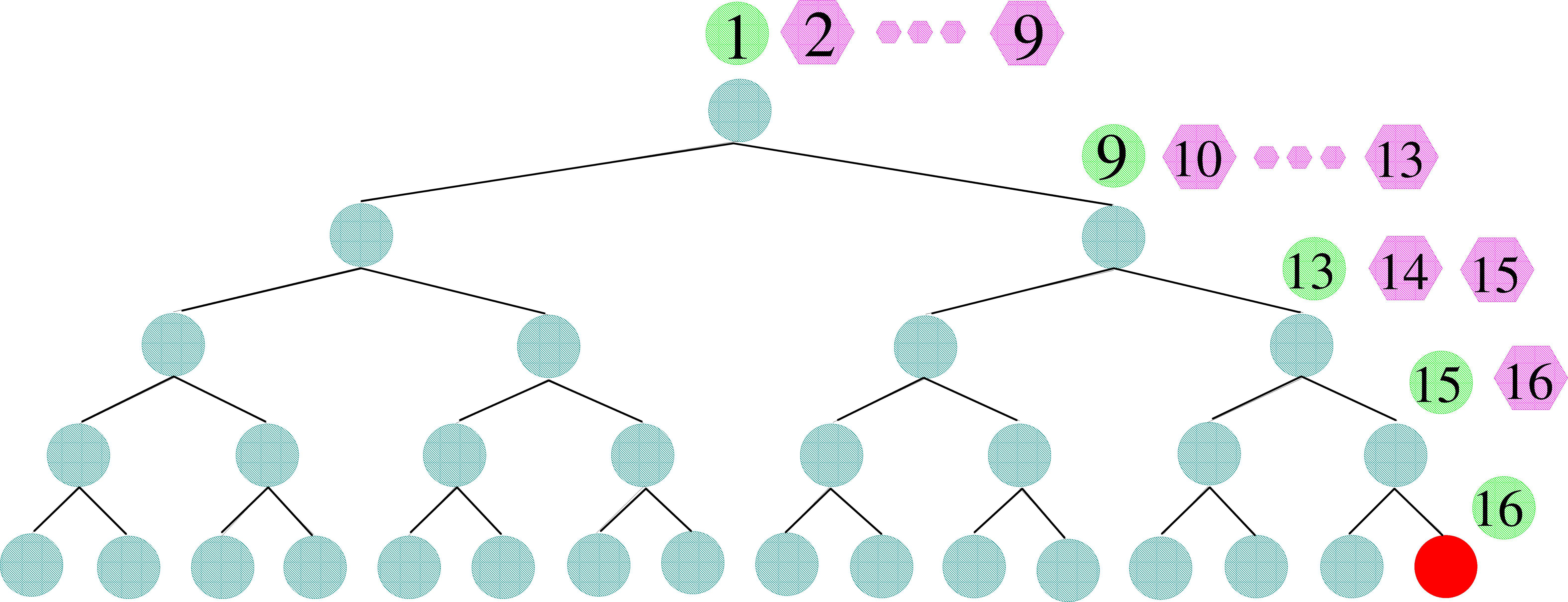}\\
	\caption{The worst case for $\STL.$}
	\label{wcstl}
\end{figure}

\noindent
Here, it is straightforward to observe that $w$ is the last node to get labelled on the network (for example, the red or black node in Figure~\ref{wcstl}).

\begin{definition}[$M_i$]
	For $i = 1,\, 2,\, \cdots, \lceil \log_{2}N\rceil,$ let $M_i(w)$ be the number of nodes $s_1, \, s_2, \, \cdots, $ having $\CID_{s_1}[i] =\CID_{s_2}[i] = \cdots = 1$ when the last node $w$ to get labelled on the network has $\CID_w[i] = 0.$
\end{definition}

\noindent Let us consider the set $\mathcal{M}(w) = \{M_0(w),\,M_1(w),\,\cdots,\,M_{\lceil \log_{2}N\rceil}(w)\}.$ First, according to the definition of the \DETNAML$(N)$ algorithm, one node gets labelled at the end of each season, so $M_i(w)$ nodes are labelled after each $M_i(w)$ seasons. As the considered node $w$ receives its label when each of the $M-1$ other nodes receives its label, we have 
\beq\label{miSomme}
\sum_{i} M_i(w) = M,
\eeq 
Second, according to the definition of $\STL(w)$, the node $w$ wakes up $M_i(w)+1$ times at any time slot $t_{2i}\in\STL(s),\, i = 1,\, 2,\, \cdots,\, \lceil \log_{2}N\rceil,$ until no other node $v$ has $\CID_v[i] = 1$ ($w$ wakes up during $M_i(w)+1$ seasons, once for the first check and $M_i(w)$ times until no other node $v$ has $\CID_v[i]=1$). 
Thus,
\beq\label{eveilSTL}
W_{\STL}(w) \leq \sum_{i}^{O(\log N)} M_i(w)+1 \leq O(M+\log N).
\eeq

\noindent
Let us now consider an arbitrary node $y.$ By definition, $\STN(y)$ is a set of even time slots during which $y$ must wake up until getting labelled to notify all nodes with a lesser $\ID$ that they are eliminated.

\begin{definition}[$L_i$]
	For $i = 1,\,2,\,\cdots,\, \lceil \log_{2}N\rceil,$ for an arbitrary unlabelled node $y$, let $L_i(y)$ be the number of unlabeled nodes with higher $\ID$s than $\ID_y$ and with $\CID[i] = 1$ when $y$ has $\CID_y[i] = 1.$
\end{definition}

For $\STL$, let us consider $\mathcal{L}(y) = \{L_0(y),\,L_1(y),\,\cdots,\,L_{O(\log N)}(y)\}.$ If $y$ has $\CID_y[i] = 1$, it wakes up at $t_{2i}\in\STN(y)$ during $L_i(y)$ seasons. As $L_i(y)$ nodes are labelled after each $L_i(y)$ seasons, we have 
\beq\label{liSomme}
\sum_{i} L_i(y) < M,
\eeq 
As a consequence,
\beq\label{eveilSTN}
W_{\STN}(y) \leq O(M).
\eeq

\noindent
Let us consider a node $z$: $z$ wakes up at only one season at any time slot $t_i\notin\{\STN(z)\bigcup\{ \STL(z)\} \}.$ Hence, we have
\beq\label{eveilOther}
W_{o}(z) \leq \sum_{i=1}^{\log N} O(1) \leq O(\log N) \leq O(\log n)\, .
\eeq
Finally, by applying (\ref{eveilSTL}), (\ref{eveilSTN}), and (\ref{eveilOther}) to (\ref{eveilTotal}), we achieve the desired result. 

\Endproof

\noindent A Maple simulation illustrates these results in Section~\ref{maple}.
\noindent
\begin{thm}\label{thm:det1}
	In single-hop beeping networks of size $n$, suppose that no node knows $n$, but a polynomial upper bound $N$ on $n$ is given in advance to all nodes and the nodes, have a unique $\ID\in \{1,\,2,\,\cdots,\, N\}.$ A deterministic naming algorithm assigns a unique label to each node in $O(n\log n)$ time slots, with no node being awake for more than $O(n \log n)$ time slots.
\end{thm}
\noindent
\textit{Proof.} By applying $M = n$ to Lemma~\ref{Lemma:detlinit}, we achieve the desired result.
\Endproof

\section{Our energy-efficient randomized algorithms}\label{namingNknown}
\subsection{The naming algorithm}
This section employs Algorithm~\ALGODETLINIT~as a subroutine protocol to design a randomized naming algorithm with $O(\log n)$ energy complexity.
\begin{definition}[Period]
	A period is the time complexity of one execution of Algorithm~\ALGODETLINIT~by $O(\log n)$ nodes and consists of $O(\log^2n)$ time slots.
\end{definition}
\begin{definition}[Group]
	A group $G_k$ is a set of nodes that will execute Algorithm~\ALGODETLINIT~during any period $k$,\\ $\left(k = 1,\, 2, \, \cdots, \,\Theta\left(\frac{n}{\log n}\right)\right).$
\end{definition}

\noindent
In what follows, we distribute the nodes into $\Theta\left(\frac{n}{\log n}\right)$ groups to obtain $\Theta(\log n)$ nodes in each group with high probability. The primary idea behind this is to make each of the $\Theta\left(\frac{n}{\log n}\right)$ groups run the \DETNAML($N$) protocol, one group after another during $\Theta\left(\frac{n}{\log n}\right)$ periods, instead of making all $n$ nodes execute it once. As each group counts at most $O(\log n)$ nodes, we have $O(\log^2 n)$ waking time per node by applying $M = O(\log n)$ to Lemma~\ref{Lemma:detlinit} for each period.\\

\noindent
Recall that we assume the total number of nodes to be unknown and the nodes to be initially indistinguishable in this section. In addition, all nodes must know a linear approximation $u$ on $n.$ This network's size approximation problem has been well studied in the distributed computing area; in particular, Brandes, Kardas, Klonowski, Paj{\k{a}}k and Wattenhofer~\cite{brandes2016approximating} designed a randomized linear size approximation algorithm, that terminated \textit{w.h.p.} in $O(\log n)$ time slots. 
\begin{enumerate}
	\item Our first idea is to make all nodes compute $u$ in $O(\log n)$ time slots using the algorithm designed by~\cite{brandes2016approximating}. This algorithm can be parameterized to make $u \in [\frac{1}{2} n,\, 2 n]$ (\textit{i.e.}, $2u \in [n, 4n]$ will be locally known by each node). Each node then sets $N = (2u)^2$ $\left(N = O\left(n^2\right)\right)$ and selects uniformly at random to be entered into one of the $\lceil\frac{2u}{\log{2u}}\rceil = \Theta\left(\frac{n}{\log n}\right)$ groups $G_1,\, G_2,\, \cdots, \,G_{\lceil\frac{2u}{\log{2u}}\rceil}.$
	\begin{lemma}\label{3logn}
          When $n$ nodes randomly and independently choose to enter $\left( \frac{n}{\log n} \right)$ groups, there are
          $\Theta(\log n)$ nodes in each group with probability $1 - O(n^{-1})$.
	\end{lemma}
	\item Each node $s$ then randomly selects a unique $\ID_s$ from $\{1,\,2,\,\cdots,\, N\}.$ 
	\item Afterwards, all groups sequentially run the \DETNAML$(N)$ protocol, one group after another:
	\begin{itemize}
		\item First, group $G_1$ nodes execute \DETNAML$(N)$ to name themselves, while all the other nodes sleep during $\lceil4 \log^2 N\rceil \footnote{Using the Chernoff Bounds, each group counts at most $2\log N$ nodes.}= \Theta(\log^2n)$ time slots. At the end of these first $\Theta(\log^2n)$ time slots, each node $s$ of group $G_1$ has a unique label $\elll(s)\in\{1,\,2,\,\cdots,\, |G_1|\}.$
		\item During some extra $\lceil\log_{2}N+1 \rceil$ time slots, the last labelled node in $G_1$ sends its label bit by bit to all nodes of the next group, $G_2.$ In parallel, each node $v$ of $G_2$ wakes up, listens to the network, and saves the received value into a variable denoted $\elll_{\prev}(v).$
		\item Second, by running the \DETNAML($N$) protocol, each node $v$ of $G_2$ receives a label $\elll(v) \in \{1,\,2,\, \cdots,\, |G_2|\}.$ Each node must update $\elll(v)\leftarrow \elll(v) + \elll_{\prev}(v)$ for its label $\elll(v)$ to belong to $ \{|G_1|+1,\,|G_1|+2,\, \cdots,\, |G_1|+|G_2|\}.$
		\item We finally make each subsequent couple of groups\\ $\{ \{G_2,\, G_3 \},\,\{G_3,\, G_4 \},\, \cdots,\,\{G_{\Theta\left(\frac{2u}{\log u}\right)-1},\,G_{\Theta\left(\frac{2u}{\log u}\right)} \}  \}$ sequentially execute these previously described computations, one couple after another.
	\end{itemize}
\end{enumerate}

To make any node $s$ know whether it has the last label of its group, we modify the \DETNAML$(N)$ protocol by making a node $s$ that was labelled during a season $S_j$ wake up during the entire season $S_{j+1}$ to listen to the network. By doing so, if $s$ hears a beep during this season, it knows that some other node will take the next label. Each node  consequently wakes up during one entire season, but these waking time slots are additive to the node's energy complexity; thus, these extra $O(\log n)$ waking time slots do not affect our $O(\log n)$ energy complexity.

\begin{thm}\label{thm:randomunknown}
	In single-hop beeping networks of size $n$, if $n$ is unknown by all nodes and the nodes are initially indistinguishable, there exists a randomized naming algorithm that assigns a unique label to each node in $O(n\log n)$ time slots \textit{w.h.p}. No node is awake for more than $O(\log^2 n)$ time slots during its execution.
\end{thm}

\noindent
\textit{Proof.}
On the one hand, the \RANDNAML($u$) algorithm uses the \DETNAML($N$) protocol as a subroutine. Let us denote the time complexity of the \DETNAML($N$) algorithm as $T_{D}.$  Per~\cite{brandes2016approximating}, $u = \Theta(n)$, and with $\Theta\left(\frac{2u}{\log{2u}}\right)$ calls of \DETNAML($N$), our \RANDNAML($u$) algorithm has

\[ 
\Theta\left(\frac{2u}{\log{2u}}\right) \times T_{D} = \Theta\left(\frac{n}{\log{n}}\right) \times T_{D} \mbox{ time complexity.}
\]
By Lemma~\ref{3logn}, the number of nodes participating in each call of \DETNAML($N$) is at most $O(\log n)$ \textit{w.h.p}. Thus, by applying $M = O(\log n)$  to Lemma~\ref{Lemma:detlinit}, we obtain $T_{D} = O(\log^2 n)$, implying $O(n\log n)$ time complexity of Algorithm~\ALGOINITN. \\

\noindent
On the other hand, each node $s$ may be awake
\begin{itemize}
	\item during one execution of the \DETNAML($N$) protocol,
	\item during $O(\log n)$ extra times to check whether it has the last label.
	\item Moreover, during $O(\log n)$ time slots when it sends its label to the next group.
\end{itemize}
\noindent
Therefore, by applying $M=O(\log n)$ to Lemma~\ref{Lemma:detlinit}, the energy complexity of \DETNAML($N$) is $O(\log^2 n).$ 
\Endproof \\

\newpage
\noindent
\underline{\textbf{Algorithm~\ALGOINITN.}}~ \RANDNAML($u$) at any node $s$\\
\framebox[1.\width]{
	\begin{algorithm}[H]
	\DontPrintSemicolon
	\SetKwInOut{Input}{Input}
	\SetKwInOut{Output}{Output}
	\Indm
	\Input{A linear approximation $u$ on $n$ computed in $O(\log n)$ time slots by executing the presented algorithm by~\cite{brandes2016approximating}.}
	\Output{The node $s$ has a unique label $\elll(s) \in \{1,\,2,\,\cdots,\,  n\}.$}
	\BlankLine
	\Indp
	$s$ sets $N \leftarrow (2u)^2$ and randomly chooses one $\ID_s$ from $\{1,\,2,\,\cdots, \,N\}.$\\
	$s$ randomly chooses to enter one group $G(s)$ from $G_1,\, G_2,\,\cdots,\, G_{\lceil \frac{2u}{\log{2u}} \rceil}$ and sets $\elll_{\prev}(s)\leftarrow0.$\\
	\For{$j$ {\upshape from } $1 \rightarrow \lceil \frac{2u}{\log{2u}} \rceil-1$}{
			
		\If{$G(s) = G_{j+1}$}{
			$s$ sleeps during $\Theta(\log^2N)$ time slots corresponding to the time complexity of the modified \DETNAML($N$) protocol executed by one group. \\
			After these $\Theta(\log^2N)$ time slots, $s$ wakes up, listens to the network during $O(\log N)$ time slots to get the last assigned label and saves this value into its $\elll_{\prev}(s).$\\
		}
		\If{$G(s) = G_{j}$}{
			$s$ sets $\elll(s)\leftarrow$\DETNAML($N$).\\
			$s$ updates $\elll(s) \leftarrow \elll(s) + \elll_{\prev}(s).$\\
			\If{$s$ is the last labeled node}{
				$s$ sends its label $\elll(s)$ bit by bit and sleeps.\\
			}
		}

	}
	
\end{algorithm}
}\par

\subsection{Adapting our naming algorithm to have a randomized counting algorithm}

\noindent Using Algorithm~\ALGOINITN, we can design a counting algorithm with $O(n\log n)$ time complexity and $O(\log^2 n)$ energy complexity on the single-hop $\BL$ network. To do so, we add the following computations after running Algorithm~\ALGOINITN~: 

\begin{enumerate}
	\item As Algorithm~\ALGOINITN~terminates after, at most, $\lceil \frac{2u}{\log{2u}} \rceil \times \lceil4 \log^2N \rceil \leq \lceil 8 n \log N \rceil$ time slots, all nodes wake up after $ \lceil8 n \log N \rceil$ time slots (counted from the first time slot of season $S_1$) to listen to the network.
	\item The last labelled node (in the last group) sends its label bit by bit. 
\end{enumerate}

\noindent
This adaptation yields the following result: 

\begin{thm}\label{thm:counting}
	In single-hop beeping networks of size $n$, if $n$ is unknown by all nodes and the nodes are initially indistinguishable, there exists a randomized counting algorithm that allows all nodes to know the exact value of $n$, in $O(n\log n)$ time slots \textit{w.h.p}, with no node being awake for more than $O(\log^2 n)$ time slots.
\end{thm}
\noindent
\textit{Proof.}
First, according to the definition of the naming problem, if, at the end of the last group's call of \DETNAML$(N)$, all nodes wake up to listen to the network, and the last labelled node sends its label bit by bit, this transmitted value corresponds \textit{w.h.p.} to the exact number of the network's nodes. 

On the other hand, as the adaptation consists of having each node wake up during extra $O(\log^2 n)$ time slots, by Lemma~\ref{Lemma:detlinit},
the time complexity of our counting algorithm is $O(n\log n)$ while its energy complexity remains $O(\log^2 n).$
\Endproof

\section{Lower bound on the energy complexity of the single-hop beeping networks naming algorithms}\label{LB}
In~\cite{chlebus2017naming}, the authors presented $\Omega(n\log n)$ lower bound for the running time
of any randomized naming algorithm. 
Note that in~\cite{chang2017exponential}, Chang \textit{et al.} studied the
energy complexity of leader election, approximate counting, and census
in several models of wireless radio networks with messages of unbounded size. In this paragraph, we conjecture a lower bound on the energy complexity of any randomized naming algorithm on radio and beeping networks.

\begin{Cnjt}\label{conj1}
	The energy complexity of any randomized algorithm that solves the naming problem
	with constant probability is $\Omega(\log n).$
\end{Cnjt}

The intuition of proof for Conjecture~\ref{conj1} lies from the proof of the lower bound on time complexity given in~\cite{chlebus2017naming}. We tried to use the Yao's minimax principle~\cite{Yao} but all our attempts in this direction failed as nodes can namely use non-uniform probabilities.

\section{Maple simulation}\label{maple}

In this section, we present a sequential Maple simulation result of Algorithm~\ALGODETLINIT's execution. This Maple simulation works as follows :
\begin{itemize}
	\item We first represent $n$ nodes by a matrix $\mathbb{M}$ with $n$ lines and $O(\log n)$ columns. Each line corresponds to the $\CID$ of a node.
	\item We then sort the matrix $\mathbb{M}$ by lines. 
	\item Afterwards, we simulate the nodes, executing Algorithm~\ALGODETLINIT~in parallel, by browsing $n$ times through $\mathbb{M}$, one column after another. Browsing once through the columns of $\mathbb{M}$ allows us to label one node and count its waking time slots number.
	\item We finally retain the maximal value of the whole nodes waking time slots number as the algorithm's empirical energy complexity.
\end{itemize}

\begin{figure}[H]
	\centering
		\includegraphics[height=7cm]{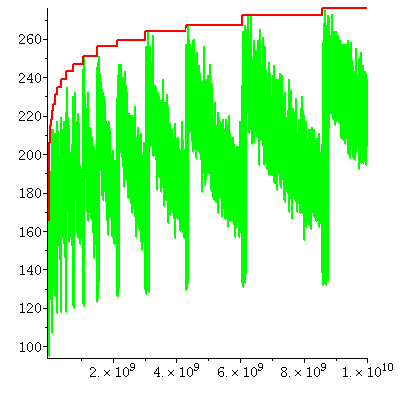}\\
		\caption{The Algorithm~\ALGODETLINIT's energy complexity for a set of $\lceil \log_{2} N \rceil$ nodes where $N = n^2$ and $n$ varies from $100$ to $10^{10}$.}
		\label{fig:Maple}
\end{figure}

\noindent
Figure~\ref{fig:Maple} compares the algorithm's empirical energy complexity (the green or grey graph) with our theoretical result (the red or black graph) for a varying number of nodes. 

\noindent
The Maple codes are available at\\ \url{https://www.irif.fr/\textasciitilde nixiton/initLoop.mw} or \\ \url{https://www.irif.fr/\textasciitilde nixiton/initLoop.pdf}

\section*{Conclusion}
In this paper, we focus on the naming problem of single-hop beeping networks of size $n$. 
When the nodes have no information about $n$ and are initially indistinguishable, we design a randomized algorithm terminating \textit{w.h.p.} in optimal $O(n\log n)$ time slots, with a $O(\log^2 n)$ energy complexity. Our algorithm can be used to solve the counting problem, returning the exact number of network nodes in $O(n\log n)$ time slots with $O(\log^2 n)$ energy complexity. 
It will be interesting to consider adapting our protocols for the multi-hop beeping network model, which is much more realistic than the single-hop model. We conjecture that $\Omega(\log n)$ is the energy complexity of any randomized naming algorithm in Beeping Networks.

%

\bibliographystyle{fundam}
\bibliography{init}

\end{document}